\documentclass[twocolumn]{jpsj2}
\usepackage[usenames]{color}
\def\runtitle{Resonant X-ray Study on the Bi-Layered Perovskite Mn Oxide
LaSr$_2$Mn$_2$O$_7$}

\def\runauthor{Y.{\sc Wakabayashi} {\it et al}.}

\title{Resonant X-ray Study on the Bi-Layered Perovskite Mn Oxide
LaSr$_2$Mn$_2$O$_7$}

\author{
Yusuke~{\sc Wakabayashi}$^{1,2}$,
Youichi~{\sc Murakami}$^{2,3,4}$\footnote{Present address : Department of Physics,Tohoku University, Sendai, 980-8578},
Ichiro~{\sc Koyama}$^2$\footnote{Present address : Department of Applied Physics, University of Tokyo, Tokyo 113-0033},
Tsuyoshi~{\sc Kimura}$^{5\dag }$,
Yoshinori~{\sc Tokura}$^{5,6}$,
Yutaka~{\sc Moritomo}$^{7}$,
Yasuo~{\sc Endoh}$^{3,8}$\footnote{Present address : Institute for Materials
Research, Tohoku University, Sendai, 980-8577}
and Kazuma~{\sc Hirota}$^{3,8}$
}

\inst{
$^1$Department of Physics, Faculty of Science and Technology, Keio University,\\
 3-14-1 Hiyoshi, Kohokuku, Yokohama 223-8522\\
$^2$Photon Factory, Institute of Materials Structure Science, High Energy 
Accelerator Research Organization, Tsukuba 305-0801\\
$^3$Core Research for Evolutional Science and Technology (CREST), Tsukuba  305-0047\\
$^4$Synchrotron Radiation Research Center, Japan Atomic Energy Research Institute (SPring-8), Mikazuki, Hyogo 679-5148\\
$^5$Joint Research Center for Atom Technology(JRCAT), Tsukuba 305-0046\\
$^6$Department of Applied Physics, University of Tokyo, Tokyo 113-0033\\
$^7$Center for Integrated Research in Science and Engineering(CIRSE), Nagoya 
University, Nagoya 464-8601\\
$^8$Department of Physics, Tohoku University, Sendai 980-8578
}

\recdate{\today}

\abst{
Charge and orbital ordering behaviors in the half doped bi-layered compound
LaSr$_2$Mn$_2$O$_7$ have been studied by resonant and non-resonant X-ray scattering.
Three different order parameters, which correspond to the A-type antiferromagnetic , a charge and an orbital ordered states, were observed by measuring the magnetostriction and the superlattice peaks characterized by wavevectors ($\frac12 \frac12 0$) and ($\frac14 \frac14 0$), respectively. 
The superlattice reflections indicating the charge and orbital ordered states were observed below 210~K. Both the intensities reach a maximum at 160~K on cooling and become very weak below 100~K. 
The peak width of the charge ordered state agrees with that of the orbital ordered state at all temperatures studied.
These results indicate that both the states originate from a single phase and that the charge/orbital ordered islands with definite interfaces disperse in the A-type antiferromagnetic phase. 
The dimensionality of the charge/orbital ordered phase is discussed using this
model. }

\kword{charge order, orbital order, X-ray scattering, layered structure}

\begin{document}
\sloppy
\maketitle

%\twocolumn

\section{Introduction\protect{\color{white}\footnote{Present address : Department of Physics, Tohoku University, Sendai, 980-8578}\footnote{Present address : Department of Applied Physics,University of Tokyo, Tokyo 113-0033}\footnote{Present address : Institute for Materials Research, Tohoku University, Sendai, 980-8577}}}
\color{black}
Among various properties which transition metal oxides exhibit, especially drastic
phenomena are the high $T_{c}$ superconductivity and the colossal magneto-resistance
(CMR). CMR has been found in some doped manganese oxides\cite{Tokura00Sci}: An
extremely large drop of the resistivity is induced by a magnetic field which is very
small in terms of the energy scale. Through recent intensive studies of CMR, it has
become recognized that fluctuations toward a orbital- and charge-ordering state and their
collapse by a magnetic field play important roles in CMR\cite{Shimomura99PRL}, which
idea is based on a homogeneous system. On the contrary, one of the recent theoretical
studies\cite{Burgy01PRL} proposes that inhomogeneities have colossal effects in
transition metal oxides and that CMR can be explained by coexistence of metal and
insulator phases. In either case, CMR and charge-ordering phenomena are considered
as results from complex interactions among charge, spin and orbital degrees of
freedom of electrons.\cite{Burgy01PRL,Brink99PRL,Mizokawa97PRB}  However, the
microscopic mechanism of CMR, orbital ordering and charge ordering are not yet fully understood.

We have been studying the stability of charge ordering as a function of the
structural dimensionality.\cite{Y.Waka,Y.Waka214} Clearly, various properties of
manganese oxide systems depend upon their structural dimensionality. For instance,
let us compare La$_{1-x}$Sr$_{x}$MnO$_3$ (LSMO113), La$_{2-2x}$Sr$_{1+2x}$Mn$_2$O$_7$
(LSMO327) and La$_{1-x}$Sr$_{1+x}$MnO$_4$ (LSMO214). Structurally, LSMO113 is an
infinite layer system thus three dimensional, and LSMO214 is a single layer system
thus two dimensional.  LSMO327 is a bi-layer system, which is in between two and
three dimensional structures. Around half doped ($x=0.5$), LSMO214 undergoes orbital ordering and charge ordering at low temperature\cite{Murakami214,Sternlieb96PRL}, while LSMO113 remains metallic\cite{Urushibara95PRB}. LSMO327, on the other hand, exhibits a phase
coexistence of a charge ordered state, an orbital ordered state and the
A-type antiferromagnetic (A-type AF) state\cite{Kubota99JPSJ}. Moreover, LSMO327 shows extremely
large CMR phenomena when the doping rate is reduced from
$x=0.5$\cite{Moritomo96Nature}, though LSMO214 remains insulating\cite{Y.Waka214}.
In the present paper, we focus on the charge and orbital ordering behaviors in the
half doped bi-layered compound LaSr$_2$Mn$_2$O$_7$, and report our comprehensive
studies using resonant and non-resonant X-ray scattering techniques.

The resonant X-ray scattering (RXS) technique is so far the only way to detect a long-range
ordered state of a local symmetry around a specific element.  This element-specific
sensitivity provides an evidence for a charge and an orbital orders. In addition,
this technique costs relatively short time (about several hours per one condition) without requiring special conditions such as a low sample temperature, thus allows us to measure under various conditions.  Although X-ray structural analysis can also clarify the local symmetry of arbitrary atom in a crystal, the data collection costs several days or more.

The known properties of the half-doped compound LaSr$_2$Mn$_2$O$_7$ are summarized
as follows: The crystal space group is $I4/mmm$ with the tetragonal lattice
constants $a=3.874$~\AA\/ and $c=19.97$~\AA\/ at room temperature. The resistivity
\cite{Kimura98PRB} becomes large between 100~K and 210~K and below 50~K, and
exhibits a large thermal hysteresis around 120~K.  The origin
of the temperature dependence between 100~K and 210~K is considered as a charge
order. In this temperature region, a transverse distortion wave with the modulation
vector ($\frac 14 \frac14 0$) was observed by electron
diffraction\cite{Kimura98PRB,Li}.  The distortion wave was attributed to an orbital order and is consistent with the CE-type bond configuration proposed by
Goodenough\cite{Goodenough} for charge ordered cubic perovskites. 
Neutron scattering experiments\cite{Kubota99JPSJ} show that the (003) magnetic scattering, which corresponds to the A-type AF order, coexists with ($\frac32 \frac32 1$) nuclear scattering, which is interpreted as the scattering from lattice distortion due to a CE-type charge order, below the N\'eel temperature $T_N^A=210$~K. 
It is also reported that a CE-type AF order emerges at $T_N^{CE}=150$~K and that this magnetic order and the charge order almost vanish below $T_L\sim 75$~K but clear CE-type magnetic scattering peaks were observed even at 15~K. 
RXS measurements by Wakabayashi {\it et al.}\cite{Y.Waka} show that the correlation length for the CE-type charge order in this compound is identical with that for the orbital order, while, in LSMO214, the correlation length for the orbital order is longer than that for the charge order\cite{Y.Waka214}. 
Argyriou {\it et al.}\cite{Argyriou} reported the phase coexistence at 125~K from a high-resolution X-ray diffraction measurement.  They also observed a lattice distortion characterized by the wavevector ($\frac 14 \frac 14 0$) and obtained the corresponding crystal structure at 160~K by structure analysis using only superlattice reflections.
 The superlattice reflections do not disappear completely even at 10~K. 
They attributed the remaining intensity at low temperature to extrinsic effects. On the other hand, Chatterji {\it et al.}\cite{Chatterji} reported the temperature dependence of the remaining scattrered X-ray intensity, which is smaller than 1~\% at 10~K compared with that at 170~K but 6 times larger than that at 50~K.  They thus proposed a reentrance of the charge order.

There still remain two problems despite the intensive studies mentioned above.
One is that the vanishing of the orbital and charge ordered phase at low temperature is not understood.
The other is that the interpretation of the observed small intensity below 50~K is controversial.
The main results of the present paper are summarized as follows:
(i) the charge order and the orbital order in this compound are two different
characters of the same phase, (ii) the structure determination of the orbital/charge ordered phase corresponding to the modulation vector ($\frac14 \frac14 0$) and ($\frac12 \frac12 0$) is demonstrated, (iii) the temperature dependence of the observed peak widths for the orbital/charge order can be understood in term of a phase segregation, and (iv) our experimental result does not show the reentrant charge order though the statistic error of our measurements is as small as that of Chatterji {\it et al.}\cite{Chatterji} and our result is consistent with high-resolution X-ray scattering study by Argyriou {\it et al.}\cite{Argyriou}.

\section{Experiment}
\subsection{Resonant X-ray scattering method}
The detection of a charge order and an orbital order is made possible by the following mechanism.
RXS is a method to detect the change in the anomalous dispersion of an ionic form factor.\cite{Murakami214,Murakami113,Nakamura99PRB,Zimmermann,Y.Waka,Y.Waka214,Ishihara1,Ishihara2}
The essence of observing an orbital state by this technique is that the local symmetry of an anisotropic form factor can be regarded as that of the corresponding valence electron orbital.
The form factor $f$ is closely related to the 1$s$-4$p_\alpha$ transition probability, where $\alpha$ is $\{x, y, z\}$.
The transition probability from 1$s$ to 4$p_\alpha$ depends on the angle between the $\alpha$-axis and the polarization vector of the incident X-rays as well as the difference between the energy of incident X-ray and the 1$s$-4$p_\alpha$ transition energy, $E-E_g^\alpha$: For example, X-rays with the polarization vector parallel to the $x$-axis produce only the $1s$-$4p_x$ transition and $E_g^y$ and $E_g^z$ do not affect the form factor for this case.
This dependence makes the form factor anisotropic when $E_g^\alpha$ is anisotropic.
An isotropic change in $E_g$ is caused by a change in the valence of a Mn ion, and an anisotropic change in $E_g^\alpha$ is caused by a change in the orbital state of a Mn ion or the surrounding crystal field.
Since $f$ is related to the $1s$-$4p$ transition, $f$ has a sharp bend at $E_g$, and it can be expressed as $f(E-E_g)$ approximately.
Destructive interference among scattering waves from ions having different $E_g$'s, $E_1$ and $E_1+\delta$, produces the amplitude $\{[f(E-E_1)-f(E-E_1-\delta)]/\delta\}\cdot \delta \simeq  \partial f / \partial E \cdot \delta$ because $\delta$ is expected to be small. 
Therefore, the bend of $f$ at $E_g$ implies that the scattering intensity from the difference in $E_g$ has a sharp enhancement at $E_g$.
The intensity from the anisotropy of a form factor depends on the rotation angle around the scattering vector called azimuthal angle $\Psi$, because the effect of $E_g^\alpha$'s on $f$ varies with the angle between the polarization vector and the $\alpha$-axis.
Therefore the azimuthal angle dependence of the intensity reflects the orbital state or the symmetry of the crystal field.

\subsection{X-ray scattering of LaSr$_2$Mn$_2$O$_7$}
X-ray diffraction measurements were performed at beamlines BL1B and BL16A2 of the Photon Factory, KEK, Tsukuba.
BL1B, which is dedicated to powder diffraction measurements, was used for measurements of the lattice constants.
This beamline is equipped with a bending magnet, an X-ray beam from which is monochromatized at the wavelength of 0.7081~\AA\/ by a flat double-crystal Si (111) monochromator and focused on the sample position by a Rh coated bent cylindrical mirror.
The spot size at the sample position is 0.5mm(vertical)$\times$0.7mm(horizontal).
X-rays scattered from a powder sample, which is typically encapsulated in a glass capillary of 300$\mu$m in diameter, are detected by a curved imaging plate. 
BL16A2 has a multi-pole wiggler as the light source, producing a linearly polarized X-ray beam. A standard four-circle diffractometer is installed there for single crystal measurements such as RXS studies.
The incident X-rays are collimated by the first mirror then monochromatized by a sagittal focusing double-crystals Si (111) monochromator which resolution is 1~eV for the incident energy near the Mn $K$-edge, 6.555~keV.
The beam is focused vertically on the sample position by the second mirror, which eliminates the higher harmonics component.
The spot size is 0.7mm(v)$\times$1.5mm(h).
An ion chamber and a scintillation counter were used as a beam monitor and a signal detector, respectively.
Closed-cycle He refrigerators were used for temperature control at both these beamlines.

The absorption correction was made for all the $E$-scan because the absorption coefficient drastically changes at $E_g\sim 6.555$~keV.
The correction was made using the absorption factors of La, Sr, O\cite{Sasaki_table} and the average of measured $f''$ for Mn$^{3+}$ and Mn$^{4+}$ ions.
The value of $f''$ for Mn$^{3+}$ and Mn$^{4+}$ are obtained by measuring the absorption spectra of LaSrMnO$_4$ and Sr$_2$MnO$_4$, respectively.
The signal intensity depends on the azimuthal angle when the form factor is anisotropic.
We define the angle as $\Psi=0^\circ$ when the $c$-axis is perpendicular to the scattering plane.
All the measurements except for the azimuthal angle scans were made at $\Psi=90^\circ$ giving a maximum intensity of RXS.

A polarization analyzer was used to measure the polarization of the scattered X-ray.
From now on, we define a $\sigma$ or $\pi$ polarization as a polarization which vector is perpendicular or parallel to the scattering plane.
The absence of the $\pi$ component in the Thomson scattering with the analyzer scattering angle $2\theta_A=90^\circ$ is used for the polarization analysis.
A Cu 220 single crystal was used for the polarization analysis because it gives a scattering angle of $2\theta_A=95.6^\circ$ when the incident photon energy $E$ is set at the Mn $K$-edge.
We define a polarimeter angle $\varphi_A$ as an angle between the scattering plane for the sample and that for the analyzer crystal.
The $\varphi_A=0^\circ$ and the $\varphi_A=90^\circ$ configurations extract the $\sigma$ polarization and the $\pi$ polarization components, respectively.

A single crystal of LaSr$_2$Mn$_2$O$_7$ was grown by the floating-zone method.
The (110) and (100) surfaces were prepared for this study.
The sample surface was polished using emery paper before each run.
Thus the sample conditions for different runs were not identical.
The typical half-width at half-maximum (HWHM) of the rocking curve was 0.1$^\circ$.
The powder sample was prepared by grinding a portion of the single crystal.

\section{Results}
\subsection{Lattice Constants}
\begin{figure}
\includegraphics[width=7cm]{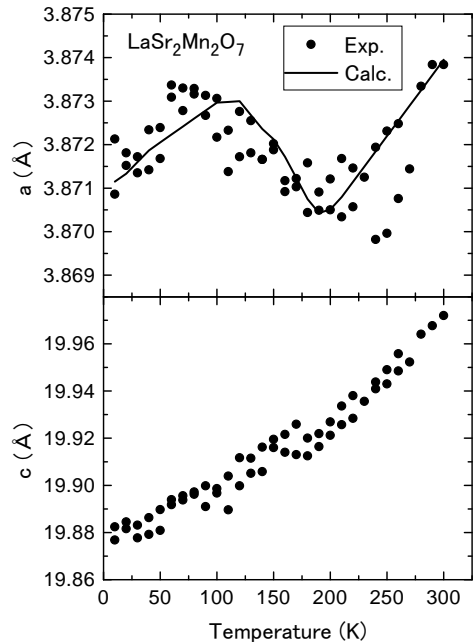}
\caption{Lattice constants $a$ and $c$ as functions of the temperature. Solid line shows the calculated value from the the A-type AF order parameter.}
 \label{fig:latcon}
\end{figure}

The lattice constants $a$ and $c$ obtained from powder X-ray diffraction measurements are shown in Fig.~\ref{fig:latcon} as functions of temperature.
They have no significant hystereses with changing temperature.
The temperature dependence of the lattice constant $c$ is dominated by a simple thermal expansion.
In contrast, the lattice constant $a$ has a characteristic ``{\it N}\/'' shape temperature dependence.
Note that the lattice constant $c$ increases 0.45\% with increasing temperature from 10~K to 300~K while the lattice constant $a$ increases only 0.09\%.
The characteristic temperature dependence of the lattice constant $a$ is brought into prominence by the small thermal expansion. This ``N'' shape is caused by magnetoelastic coupling, which we will discuss later in detail.
The temperature dependence of the lattice constants is extremely different from that of the superlattice intensity.

\subsection{Temperature Dependence of Superlattice Reflections}
\begin{figure}
%profiles.eps

%label\{fig:profiles\}

%\special{epsfile=profiles.eps hsize=240}
\includegraphics[width=7cm]{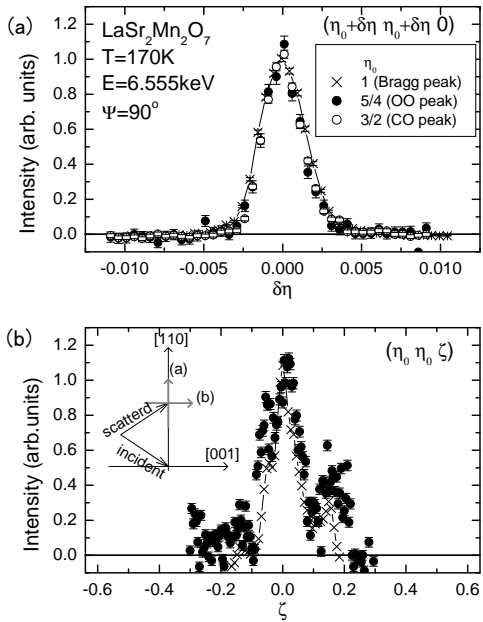}
%\vspace{12cm}
\caption{Peak profiles of (110), ($\frac32 \frac32 0$) and ($\frac54 \frac54 0$) reflections at  170~K with $E=$6.555~keV along (a) the [110] direction and (b) the [001] direction. The profile for the [001]-direction consists of three peaks, reflecting that the sample is composed of three domains.
}
\label{fig:profiles}
\end{figure}

As described in Sec.~1, there appear two types of superlattice reflections at low temperatures, namely the ($\frac 14 \frac14 0$) and ($\frac12 \frac12 0$) type reflections, which correspond to the orbital ordering (OO) and charge ordering (CO) in the CE type model, respectively.
Both types of superlattice reflections were observed by RXS at $E=6.555$~keV in the temperature region between 100~K and 210~K.
Typical results of longitudinal and transverse scans at 170~K are shown in Fig.~\ref{fig:profiles}.
The widths for the [110] and the [001] directions of the superlattice reflections are nearly equal to those of the Bragg reflections, which are limited by the instrumental resolution and the mosaicity of the crystal: the resolution limit of this experiment was about 2500\AA.
The profile along the [001] direction consists of a main peak and two small peaks, which indicates that the sample has three domains in the beam spot area. Profiles along [001], i.e., $\omega$-scan profiles, varied slightly in different spots and different runs. We thus mainly focused on $\omega$-$2\theta$ scans, which are free from domain distributions, for measuring the peak widths in the present study.

\begin{figure}
\includegraphics[width=8cm]{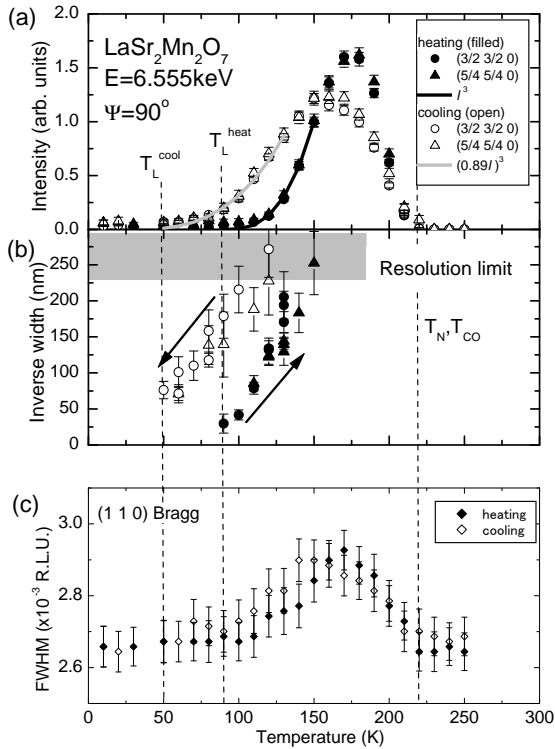}
\caption{(a) Temperature dependence of the intensities of the ($\frac32 \frac32 0$) and ($\frac54 \frac54 0$) superlattice reflections with $E$=6.555~keV. The ($\frac54 \frac54 0$) intensity is multiplied by 3.2 so as to be directly compared with the ($\frac32 \frac32 0$) intensity. The superlattice reflections are enhanced between $T_L$ and $T_N$.
(b) Temperature dependence of the inverse peak widths, i.e., the correlation lengths $\xi$ or domain sizes $l$, for the charge order(circles) and the orbital order(triangles) calculated from the peak widths of the ($\frac32 \frac32 0$) and ($\frac54 \frac54 0$) superlattice reflections. 
(c) Temperature dependence of the peak width of the (110) Bragg reflection. The width is broad between $T_L$ and $T_N$. R.L.U. stands for reciprocal lattice unit.}
 \label{fig:super-temp}
\end{figure}

We have focused upon ($\frac54 \frac54 0$) and ($\frac32 \frac32 0$) for the ($\frac14 \frac14 0$)-type OO peak and the ($\frac12 \frac12 0$) type CO peak.
Temperature dependence of the two peaks is shown in Fig.~\ref{fig:super-temp}(a).
Note that the ($\frac54 \frac54 0$) intensity is multiplied by 3.2 so as to be directly compared with the ($\frac32 \frac32 0$) intensity.
Both the intensities have a similar temperature dependence and a large hysteresis around 120~K.
Both the superlattice reflections appear at 210~K on cooling and reach a maximum at 150~K, then decrease with further cooling.
The maximum was observed at 170~K on heating.
At 10~K, both the intensities decrease to approximately 1\% of the maximum values though the peak widths are as narrow as that of the Bragg reflection.
These remanent peaks correspond to the OO and CO phase because they completely disappear above 250~K.
As for ($\frac 54 \frac54 0$), the intensity and width show no temperature dependence below 90~K for heating and 50~K for cooling.  This result is consistent with ref.\citen{Argyriou}, which reported no temperature dependence of the ($\frac 74 \frac94 0$) intensity in this temperature range.  However, ref.~\citen{Chatterji} reported that the ($\frac54 \frac34 0$) intensity is temperature dependent in this region.  We thus studied ($\frac54 \frac34 0$) and confirmed that the temperature dependence is similar to that of ($\frac54 \frac54 0$).  The origin of this reflection will be discussed later.

To extract the correlation lengths of the OO and CO peaks, we subtracted the remanent intensities at 10~K from the data below 150~K.
The peak profiles were fitted with a Gaussian function convoluted with the instrumental resolution. 
Thus obtained peak widths are shown in Fig.~\ref{fig:super-temp}(b) in the form of the inverse width, which corresponds to the domain size $l$ or the correlation length $\xi$ along the [110] direction.
As shown in Figs.~\ref{fig:super-temp}(a) and (b), both the ($\frac 54 \frac54 0$) and ($\frac 32 \frac32 0$) peaks have similar temperature dependence of the intensities and widths.
The width of the (110) Bragg reflection is shown in Fig.~\ref{fig:super-temp}(c) as a function of temperature.
The width was found to become broad in the temperature region in which the superlattice reflections are enhanced.

\subsection{Energy Dependence of Superlattice Reflections}\label{sec.str}
\begin{figure}
\includegraphics[width=7cm]{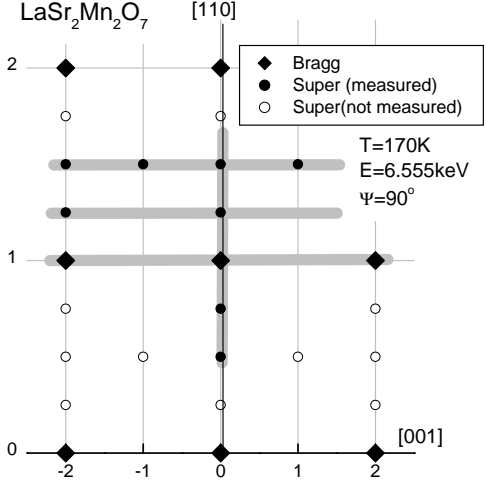}

\caption{Schematic view of the reciprocal space at 170~K with $E$=6.555~keV. Closed rectangles and closed circles represent the observed Bragg and superlattice reflections, respectively, and the open circles represent the position where the superlattice reflections are expected to be observed. The measured region is shown by the gray line.}
 \label{fig:recipro}
\end{figure}

\begin{figure}
\includegraphics[width=8cm]{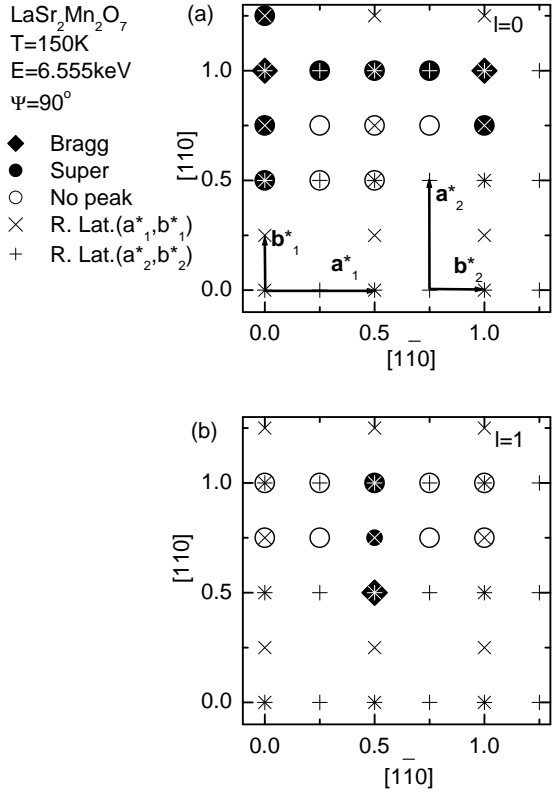}
\caption{Schematic view of the reciprocal space at 150~K with $E$=6.555~keV for (a) $hk0$-plane and (b) $hk1$-plane. Closed diamonds and closed circles denote the observed Bragg and superlattice reflection, respectively, and the open circles denote the position where the no intensity was observed. These peaks form $2\sqrt2 \times \sqrt2 \times 1$ ($\mib a_1$*-$\mib b_1$*) and its twin, $\sqrt2 \times 2\sqrt2 \times 1$ ($\mib a_2$*-$\mib b_2$*), reciprocal lattice with systematic extinction $h+l=2n$, which indicates a $B$-base centered lattice. The reciprocal lattice points for both twins are denoted by the symbols $\times$ and $+$.}
\label{fig:recipro-e}
\end{figure}

\begin{figure}
\includegraphics[width=8cm]{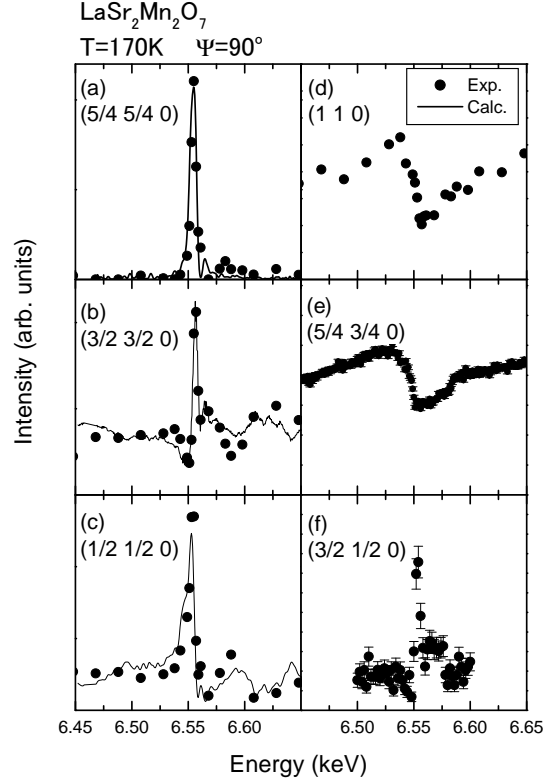}
\caption{Energy dependence of the integrated intensity of (a) ($\frac54 \frac54 0$), (b) ($\frac32 \frac32 0$), (c) ($\frac12 \frac12 0$), (d) (110), (e) ($\frac54 \frac34 0$) and (f) ($\frac32 \frac12 0$) reflections at 170~K with $\Psi=90^\circ$. The absorption correction was made using the average of measured $f''$ for Mn$^{3+}$ and Mn$^{4+}$ ions and absorption factor of La, Sr, O\protect{\cite{Sasaki_table}}. Calculated intensity is also shown by solid lines in the panels (a)-(c).}
\label{fig:energy}
\end{figure}

To determine the unit cell of the CO/OO phase, we searched for superlattice reflections in the ($hhl$), ($hk0$) and ($hk1$) planes at 170~K. 
The results are summarized in Figs.~\ref{fig:recipro} and \ref{fig:recipro-e}.
The results indicate the unit cell of this phase is $\mib a-\mib b$, $2\mib a+2\mib b$, and $\mib c$, where $\mib a$, $\mib b$ and $\mib c$ are the basis vectors of the structure at room temperature. 
Since the crystal system at room temperature is tetragonal, this $\sqrt 2 \times 2\sqrt 2 \times 1$ structure is twinned with the $2\sqrt 2 \times \sqrt 2 \times 1$ structure.
The reciprocal lattice points for the intrinsic two domains are illustrated in Fig.~\ref{fig:recipro-e} as crosses and plus signs.
The peak position shows the systematic extinction $h+l=2n$, which indicates a $B$-base centered lattice.
Figure~\ref{fig:energy} shows the $E$-dependence of the intensities of selected superlattice reflections.
As expected for the CO and OO phase, the peak intensities of the superlattice reflections at ($\frac 54 \frac54 0$), ($\frac 12 \frac12 0$) and ($\frac 32 \frac32 0$) are enhanced near the Mn $K$-edge.
However, the ($\frac54 \frac34 0$) reflection, which was ascribed to a reentrant charge ordering in ref.~\citen{Chatterji}, shows no enhancement.
This lack of the enhancement and similarity to the (110) energy dependence denote that this peak is mainly caused by displacements of Mn ions.
We have carried out some model calculations of the energy dependence using the values of $f''$ experimentally determined from LaSrMnO$_4$ and $f'$ obtained by the Kramers-Kronig transformation of $f''$. 
We found that the energy dependence of the ($\frac 54 \frac54 0$) intensity is proportional to $|\partial f / \partial E|^2$ as shown in Fig.~\ref{fig:energy}(a).
On the other hand, as shown in Figs.~\ref{fig:energy}(b) and (c), the ($\frac12 \frac12 0$) and ($\frac32 \frac32 0$) intensities are described as $|\partial f / \partial E + C|^2$ where $C$ is an amplitude of a real number.
The finite $C$ is ascribed to the displacement of oxygen induced by the valence and orbital states of the Mn ions.
More detailed analyses on these results are described in \S~\ref{sec:e-dep}.

\subsection{Polarization and Azimuthal Dependence of Superlattice Reflections}
\begin{figure}
%pol.eps

\includegraphics[width=8cm]{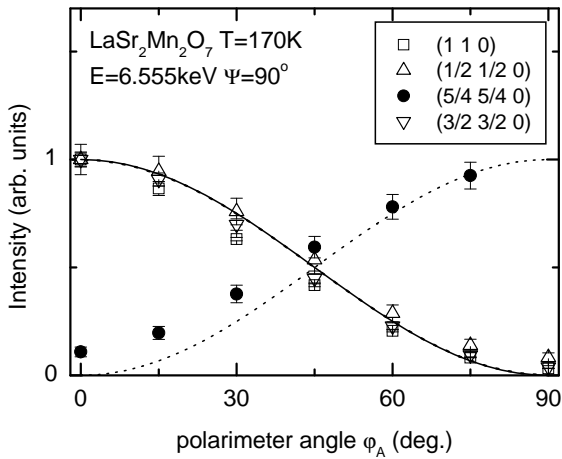}
\caption{Polarimeter angle dependence of the Bragg and the superlattice intensity at 170~K with $\Psi=90^\circ$. Solid line and dotted line show the calculated values for $\sigma$- and $\pi$ polarization, respectively. The maximum value of each data set is normalized to unity.}
\label{fig:pol}
\end{figure}

\begin{figure}
\includegraphics[width=7cm]{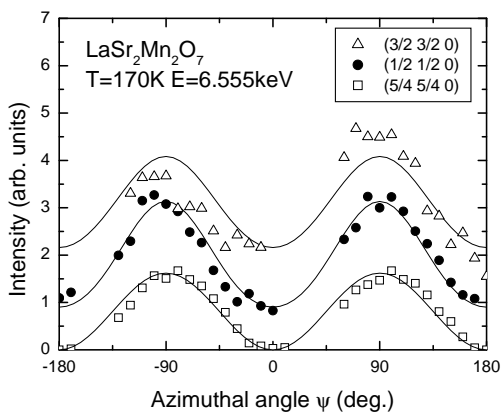}
\caption{Azimuthal angle dependence of the intensity of several superlattice reflections. The solid lines for ($\frac32 \frac32 0$) and ($\frac12 \frac12 0$) show the results of the fitting to $|A+B\sin \Psi|^2$ and that for ($\frac54 \frac54 0$) the fitting to $|A \sin \Psi|^2$. }
 \label{fig:azi}
\end{figure}

\begin{figure}
%aziene.eps

%label\{fig:aziene\}

%\special{epsfile=aziene.eps hsize=240}
\includegraphics[width=7cm]{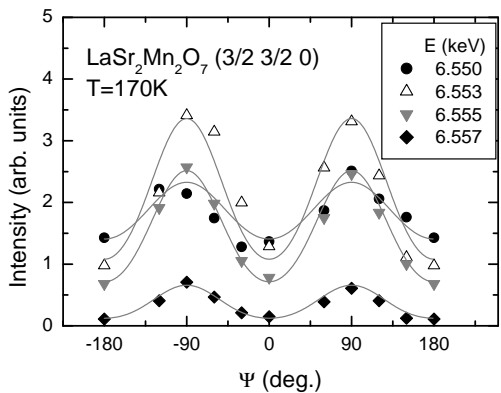}
%\vspace{6cm}
\caption{The intensity at ($\frac32 \frac32 0$) superlattice reflection for several $E$'s. The solid curves for each plots show the results of the fitting to $|A+B\sin \Psi|^2$.}
\label{fig:aziene}
\end{figure}

We measured the polarization dependence of selected ($hh0$) reflections. 
Figure~\ref{fig:pol} shows the results as a function of the polarimeter angle $\varphi_A$.
The solid and dotted lines in the figure denote calculated intensities for the $\sigma$- and the $\pi$ polarizations, which indicate that the ($\frac 14 \frac14 0$)-type reflection (OO) has the $\pi$ polarization while the ($\frac 12 \frac12 0$)-type reflection (CO) has the $\sigma$ polarization as Bragg reflections.
The rotation of the polarization shown in this figure is caused by the anisotropy of a scattering factor.
The anisotropic scattering factor also makes the azimuthal angle dependence of the scattering intensity.
The azimuthal angle dependence of the ($hh0$) superlattice reflections is shown in Fig.~\ref{fig:azi}.
The intensity of the ($\frac 12\frac 12 0$)-type reflections is normalized by the intensity at an off-resonant energy ($E=6.45$~keV) to make an accurate correction for the sample shape.
The ($\frac54\frac54 0$) intensity is normalized by the (110) Bragg intensity because the ($\frac 54 \frac54 0$) reflection has no intensity at off-resonant energies.
The azimuthal dependence of the ($\frac54 \frac54 0$) and ($\frac12 \frac12 0$)-type intensities can be fitted to $\sin ^2 \Psi$ and $|A+B\sin \Psi|^2$ ($A$ and $B$ are real numbers), respectively.
Azimuthal angle dependence of the ($\frac 32 \frac 32 0$) at selected energies (four different energies) is shown in Fig.~\ref{fig:aziene}.
These results have the intensity maxima at $\Psi=\pm 90^\circ$.
This result indicates that changing the incident energy causes little change in this dependence.

\section{Discussion}
\subsection{Structure of the Orbital Order}\label{sec:e-dep}
In this section, we demonstrate the structure determination of the orbital order, refering to the extinction rule and the azimuthal angle dependence of the intensity.
The peak widths of the superlattice reflections in the $c$*-direction at 170~K are nearly equal to the widths of the Bragg reflections, which are limited by the instrumental resolution and the mosaicity of the crystal, as shown in Fig.~\ref{fig:profiles}.
This result indicates that the charge and orbital orders have long range correlations in the $c$-direction.
It gives contrast to the result for the single layer compound La$_{0.5}$Sr$_{1.5}$MnO$_4$\cite{Y.Waka214}, which shows a short range correlation along the $c$-direction.
This suggests that the magnitude of the inter-layer interaction of the bi-layered compound is different from that of the single-layered compound.
The inter-layer interaction is closely related to the structure of the orbital order in the $c$-direction.

As mentioned above, the unit cell of the orbital ordered phase has the dimensions $\sqrt{2}$$a$$\times$$2\sqrt{2}$$a$$\times$$c$.
As the origin of the anisotropy of the scattering factor, two mechanisms have been proposed; the Coulomb mechanism\cite{Ishihara1,Ishihara2} and the JT mechanism\cite{Elfimov,Benfatto,Takahashi}.
In the former, the anisotropy of the form factor is attributable to the Coulomb interaction between $3d$ and $4p$ electrons.
The latter, on the other hand, the anisotropy is regarded as the band effect due to the lattice distortion, which is caused by the anisotropy of the $e_g$-electron density through the Jahn-Teller effect.
In either case the anisotropy of the scattering factor is considered to reflect the anisotropy of the $e_g$-electron density due to the orbital ordering.

\begin{figure}
%label\{fig:Str-model\}

%\special{epsfile=models.eps hsize=240}
%\vspace{10cm}
\includegraphics[width=7cm]{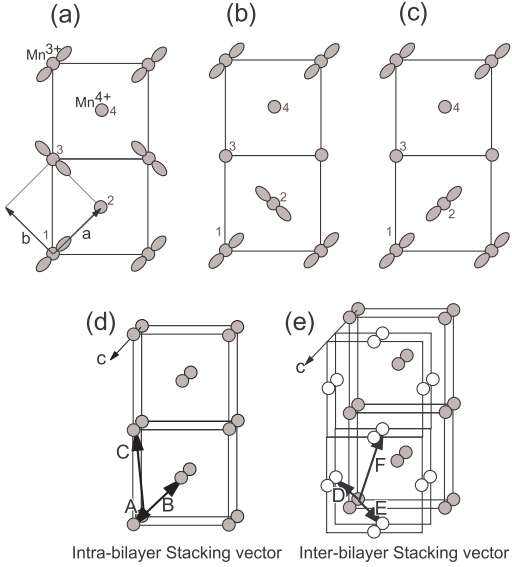}
\caption{(a)-(c) The possible arrangements of the orbital order in a plane and (d),(e) the possible stacking vectors. The gray and white circles denote the position of Mn ions and the ellipsoidal symbols denote the $e_g$ electrons. The resulting structure is the (a)-type with stacking vectors A and E.}
\label{fig:Str-model}
\end{figure}

We assume that each layer in a double MnO$_2$ sheet has the same orbital structure and the whole structure is formed by stacking of such layers with  certain intra- and inter- bilayer stacking vectors.
This assumption is reasonable because the MnO$_2$ planes are crystallographically equivalent at room temperature and there is no superstructure in the $c$-direction.
Based on this assumption, we will consider the orbital structure, that is, the spatial arrangement of Mn$^{4+}$ and two kinds of Mn$^{3+}$ (Mn$^{3+}_A$ and Mn$^{3+}_B$).
Possible structures in a MnO$_2$ plane having an area of $\sqrt 2 a \times 2\sqrt 2 a$ are shown in Figs.~\ref{fig:Str-model}(a)-(c).
The polarization property of the superlattice reflection shows that the Mn$^{3+}$ site is on a reflection plane perpendicular to the $c$-axis; The details of this assignment will be shown in the next section.
This symmetry requires that the form factors for Mn$^{3+}_A$ and Mn$^{3+}_B$ are equal at $\Psi=0^\circ$.
The structure factor of the ($\frac 14 \frac14 0$)-type reflection is proportional to $f_1+if_2-f_3-if_4$, where $f_i$ denotes the form factor of Mn ion numbered $i$ in Fig.~\ref{fig:Str-model}(a).
For the structure (a), the structure factor is porportional to the difference between the form factor of Mn$^{3+}_A$ and that of Mn$^{3+}_B$.
The intensity of ($\frac 54 \frac54 0$) at $\Psi=0^\circ$ is zero as shown in Fig.\ref{fig:azi}.
This result indicates the structure factor of the ($\frac 14 \frac14 0$)-type reflection is zero at $\Psi=0^\circ$, therefore the structure in the plane is (a).
Unlike this, for the structures (b) and (c), the structure factor of ($\frac 14 \frac14 0$)-type reflection contains the difference between the form factor of Mn$^{3+}$ and that of Mn$^{4+}$.
There is no condition in which the form factor of Mn$^{3+}$ equals that of Mn$^{4+}$.

Next, let us consider the stacking structure, which consists of an intra-bilayer stacking and an inter-bilayer stacking.
The possible intra-bilayer stacking vectors are (A) (0,0,0.18), (B) (1,0,0.18) and (C) (1,1,0.18), which are shown in Fig. \ref{fig:Str-model}(d), and those for the inter-bilayer stacking are (D) ($\frac12$,$\frac12$,$\frac12$), (E) ($\frac12$,$\bar{\frac{1}{2}}$,$\frac12$) and (F) ($\frac32$,$\frac12$,$\frac12$), which are shown in Fig. \ref{fig:Str-model}(e).
We have concluded that the intra-bilayer stacking vector is (A), because the (B)- and the (C)-stacking cause the absence of ($\frac{2n+1}{2}$$\frac{2n+1}{2}$0) and ($\frac{2n+1}{4}$$\frac{2n+1}{4}$0) reflections, respectively.
The inter-bilayer stacking vector is (E), because, as mentioned in \S~\ref{sec.str}, the Bravais lattice of this compound is $B$-centered orthorhombic. 
More specificly, (D)-stacking makes the ($\frac54 \frac54 1$) reflection, which were not observed, and (F)-stacking results in the absence of ($\frac14 \frac14 0$).
Consequently, it was found that the orbital structure of this compound is Fig.~\ref{fig:Str-model}(a)-type structure with the stacking vectors (A) and (E).
The intra-bilayer stacking is same as the stacking of cubic perovskite compounds with the CE-type order, such as Nd$_{1/2}$Sr$_{1/2}$MnO$_3$ and Pr$_{1/2}$Ca$_{1/2}$MnO$_3$.
This implies the cubic- and bi-layered perovskite compounds have a common interaction which induces the orbital order.

The $E$-dependence of the intensity shown in Fig.~\ref{fig:energy} can be explained from the stacking vectors (A) and (E).
The structure factors for this $\sqrt2 \times 2\sqrt2 \times1$ structure at ($\frac12 \frac12 0$) and ($\frac32 \frac32 0$) with $\Psi=90^\circ$ are
\begin{eqnarray}
F\left(\frac12 \frac12 0\right)&=&
 8(f^{3+}-f^{4+})-32f_{\rm O} \sin\left(\pi\delta/2\right),\label{eq:COstrF12}\\ 
F\left(\frac32 \frac32 0\right)&=&
 8(f^{3+}-f^{4+})+32f_{\rm O} \sin\left(3\pi\delta/2\right),\label{eq:COstrF32}
\end{eqnarray}
where $f^{n+}$ and $f_{\rm O}$ are the form factor of the Mn$^{n+}$ and the oxygen, respectively, and $\delta$ is the breathing mode displacement of the oxygen atoms.
The solid lines in Figs.\ref{fig:energy}(a) and (b) are the results of the fitting to these equations.
The experimental results are well reproduced.

\subsection{Anisotropic Scattering Factor Tensor at 170~K}
The element of the scattering factor tensor $f_{\alpha \beta}$, where $\alpha$ and $\beta$ are ${x, y,}$ or ${z}$, is brought about by the process that an electron excites from 1$s$ to 4$p_\alpha$ and relaxes from 4$p_\beta$ to 1$s$.
In the ATS scattering, we have to deal with an ellipsoid of scattering function, which has three principle axes in the reciprocal space.
Thus, when the Mn site has an inversion symmetry, one can chose a certain coordinate to make the scattering factor tensor a diagonal tensor. 
In bi-layered compounds, MnO$_6$ is almost a regular octahedron so that the form factor tensor is approximately diagonal.
A regular octahedron has three fourfold rotation axes, six twofold rotation axes and four threefold rotation axes.
A principal axis of the form factor is expected to coincide with one of them.
If we choose a threefold axis as a principal axis of the tensor, the scattering with $\pi$ polarization is not expected on the [110]*-axis for any orbital structure. This is the reason for the existence of the reflection plane on the Mn$^{3+}$ site which was used in the previous section.
Thus, the principal axes are (i) the three fourfold rotation axes $x$, $y$ and $z$ , or (ii) a fourfold rotation axis $x'$ and the two twofold rotation axes $y'$ and $z'$, which are perpendicular to the $x'$ axis.
We denote the principal value of the scattering factor tensor for the principal axis $\alpha$ as $f_\alpha$.
The $\sigma$ incident, $\pi$ scattered element of the structure factor tensor at ($\frac54 \frac54 0$) for the set (i) is $(f_y-f_x)\sin\Psi\cos\theta$ and for the set (ii) is $\frac12 (f_{y'}+f_{z'}-2f_{x'})\sin\Psi\cos\theta+\frac{1}{\sqrt{2}}(f_{y'}-f_{z'})\cos(2\Psi)\sin\theta$.
The experimental result shows that the intensity is proportional to $\sin^2 \Psi$, which agrees with the expectation from the set (i).
Therefore, we choose the set (i) as the principal axes of the scattering factor tensor for this compound.

Next, we discuss the out-of-plane anisotropy of the scattering factor tensor.
As shown in Figs.\ref{fig:azi} and \ref{fig:aziene}, the ($\frac n2 \frac n2 0$) intensity has maxima at $\Psi_{\mbox{\small max}}=\pm 90^\circ$, which are independent of $n$ and $E$.
The structure factors for the $\sigma$ polarization are
\begin{eqnarray}
F^{\sigma}\left(\frac12 \frac12 0\right)
&=&\!\!\!\!8(f_z-f^{4+})+4(f_x+f_y-2f_z)\sin ^2 \Psi \nonumber\\
&&- 32f_{\rm O}\sin(\pi \delta /2) ,\label{eq:FCO}\\
F^{\sigma}\left(\frac32 \frac32 0\right)
&=&\!\!\!\!8(f_z-f^{4+})+4(f_x+f_y-2f_z)\sin ^2 \Psi \nonumber\\
&&+ 32f_{\rm O}\sin(3\pi \delta /2) ,\label{eq:FCO2}
\end{eqnarray}
and those for the $\pi$ polarization are small.\cite{Ishihara2}
These structure factors are same as eq.(\ref{eq:COstrF12}) and eq.(\ref{eq:COstrF32}) when we substitute $(f_x+f_y)/2$ for $f^{3+}$ and $\Psi$ for 90$^\circ$.
Let us recall that the $f_\alpha$ can be written in the form $f(E-E_g^\alpha)$ approximately.
It suggests that the azimuthal angle dependence should be explained by the anisotropy of the absorption-edge energy.
To reproduce the experimental results, the value of $E_g^z$ has to be larger than the average of $E_g^x$ and $E_g^y$.
A possible origin of the anisotropy of $E^\alpha$ is not only the effect of the $e_g$ electron, such as Coulomb interaction or Jahn-Teller distortion, but also the layered structure\cite{Ishihara_layer}.
In fact, it is known that the EXAFS spectra for some layered compounds are highly anisotropic.\cite{Merz}
Although the origin could not be determined from this experiment, a theoretical analysis shows the effect of the layered structure is dominant.\cite{Ishihara_layer}

\subsection{Temperature dependence of the lattice constants}
The unusual temperature dependence of the $a$-lattice constant can be understood in terms of the magnetoelastic coupling corresponding to the A-type AF order.
The $t_{2g}$ spins, which have a greater part of the moment are localized.
The magnetoelastic effect of localized moments causes magnetostriction proportional to the square of the ordered moment size,\cite{Lee,Waka_DyZn2}
which is proportional to the intensity of the neutron magnetic scattering.
Thus, the change of the lattice constant at $T_N$ is expected to be proportional to the magnetic scattering intensity corresponding to the A-type AF order.
For this compound, the result of the magnetic scattering measurement was reported in ref.\citen{Kubota99JPSJ} and we use it for our analysis.
The calculated $a$-lattice constant is shown in Fig.~\ref{fig:latcon} as a solid line.
The non-magnetic lattice constant $a_0$ at temperature $T$ is assumed to be $a_0(T)=\beta T+a_0(0)$, where $\beta$ and $a_0(0)$ are obtained from the thermal expansion above $T_N$.
%The reasons for using such $a_0$ are the absence of the certain reference material, i.e., a material having the same structure with no magnetism, and expected low Debye temperature from the temperature dependence of the $c$-lattice constant.
The magnitude of the magnetoelastic effect was treated as a fitting parameter.
The calculated value shows a good agreement with the experimental results.
As a consequence, the temperature dependence of the lattice constant is understood by taking into account only the effect of the A-type AF order.
Neglecting the CE-AF for above discussion is valid because the result of the neutron magnetic scattering indicates that the fraction of the A-type AF order is five-times as large as that of the CE-type AF order\cite{Kubota99JPSJ}.
The temperature dependence of the lattice constant is very different from that of the intensity of the superlattice reflection.
This difference means the existence of more than two different order parameters and suggests a phase coexistence.
The broadening of the (110) Bragg reflection around 150~K shown in the inset of Fig.~\ref{fig:super-temp}(b) also supports the occurrence of the phase coexistence.

\subsection{Temperature Variation of the Charge Order and the Orbital Order}
The intensity of the superlattice reflection is proportional to the volume fraction of corresponding phase and the inverse of the peak width gives the domain size $l$ or correlation length $\xi$.
Here, the order parameters of the charge and orbital orders are expected to be constant because other half-doped cubic perovskite manganites show sharp first order charge/orbital ordering transition.\cite{Nakamura99PRB,Zimmermann}
The order in the cubic perovskites are similar to that in LSMO327 in terms of its stacking structure or correlation length.
The correlation length of the charge ordered phase in Pr$_{0.6}$Ca$_{0.4}$MnO$_3$ diverges toward the ordered temperature\cite{Zimmermann}: this behavior is very similar to that of the LSMO327 around 210~K. 
The charge order transition of the half-doped LSMO214 is not a sharp one, but the order in this compound is different from that in LSMO327 in term of correlation length.
In LSMO214, the correlation is always short ranged.

The temperature region with enhanced superlattice reflections coincides with that for a large resistivity\cite{Kimura98PRB}, hence the excess resistivity is ascribed to the charge order.
The peak width in the charge ordered phase agrees with that in the orbital ordered phase at all temperatures studied, though these two widths are not necessarily equal: Different peak widths for these orders were observed in Pr$_{1-x}$Ca$_{x}$MnO$_3$\cite{Zimmermann} and LSMO214\cite{Y.Waka}, whose charge order and the orbital order structures are same as that of LSMO327.
The present result can be explained by two different models; That is, (A) $\xi$ of the charge ordered phase coincides with that of the orbital ordered phase, or, (B) the peak width reflects the size of the ordered domain which consists of the charge {\it and\/} orbital ordered phase.
The model (A) implies that such accidental coincidence is realized at all the temperatures measured, which is implausible.
The model (B) does not have such an unreasonable assumption.
This model implies that a domain which is orbitally ordered is also chargely ordered, i.e., the charge order and the orbital order in this compound originate from a single phase; we thus call this phase the charge/orbital ordered phase.
Therefore, in this compound, there are only two phases though it seems that there are three phases (charge ordered phase, orbital ordered phase and the A-type phase) at first sight.
This model also implies that islands of the charge/orbital ordered phase with definite interfaces disperse in the A-type phase.
This phase coexistence is consistent with the above results showing two different order parameters, which correspond to the A-type and CE-type, defined by the magnetostriction and superlattice peaks respectively.

Next, we consider how the charge/orbital ordered islands grow or melt in the A-AF phase below 150~K.
As mentioned above, $l$ corresponds to the domain size in this system.
Thus, it is expected that the intensity of the superlattice reflection, which is the volume of the charge/orbital ordered state, is proportional to $Nl^{D}$ where $N$ is the number of the ordered islands and $D$ denotes the dimensionality of the growing or melting.
If we assume that $N$ is conserved as temperature varies, the volume fraction of the ordered phase is proportional to $l^D$.
Results of fitting the temperature dependence of the superlattice intensity with $(A_hl_h)^3$ or $(A_cl_c)^3$, where $A_h$ and $A_c$ are the scaling parameters, are shown in Fig.~\ref{fig:super-temp} with the solid and dashed lines, respectively, 
The fitting results are very good.
We tried fitting with $D_{h(c)}=2$ and 4, but they produce worse agreement.
Thus the experimental result indicates that the value of $D$ is three in the melting and growing processes of the ordered phase between 100~K and 150~K; The result $D=3$ coincides with the expectation from the Landau free energy with a surface tension term.
$A_h$ can be normalized to unity because the intensity is represented in arbitrarily units, and the normalized $A_c$ is $0.89$.
The small deviation of $A_c$ from unity indicates that almost all the domains are not divided by cooling.
This deviation is ascribable to the correlated small charge/orbital ordered domains which were made by dividing a large domain on cooling.
%This melting behavior indicates that the charge/orbital order is , because the existence of $n$ such nuclei in a large domain makes divided $n$ domains centered the nuclei when the domain shrinks.

\section{Summary}
Resonant X-ray scattering measurements have been performed on LaSr$_2$Mn$_2$O$_7$ to elucidate the charge and the orbital ordering.
We have observed anisotropic form factor of Mn ion near the $K$-absorption edge.
The structure determination of the charge/orbital order was demonstrated using the extinction rule and the anisotropy of the form factor.
The anisotropy of the form factor indicates that the energy level of $4p_z$ is higher than that of in-plane 4$p_x$ and 4$p_y$.
This electronic structure is attributable to the orbital ordering of $e_g$ electrons and the layered structure of the compound.
The temperature dependence of the lattice constant is ascribed to the magnetoelastic effect in the A-type AF phase while the superlattice reflections come from a long-range order of charge and orbital in the CE-type AF phase.
The correlation length for the charge order is same as that for the orbital order, while the two lengths are, in general, not related.
The results indicate the coexistence of the A-type AF phase and CE-type AF phase.
Moreover, the growth and melting of the charge/orbital order were observed, showing the dimensionality of the order is three dimensional.

\section*{Acknowledgement}
The authors thank Dr.\ S.~Ishihara,  Prof.\ S.~Maekawa, Prof.\ N.~Wakabayashi, Dr.\ Doon Gibbs and Dr.\ J.~P.~Hill for valuable discussions.  
This work was supported by a Grant-In-Aid for Scientific Research from the Ministry of Education, Science and Culture, Japan, by the New Energy and Industrial Technology Development Organization (NEDO), the U.S.DOE under Contract No. DE-AC02-98CH10886, and Core Research for Evolutional Science and Technology (CREST).


\begin{thebibliography}{99}
\bibitem{Tokura00Sci} Y.~Tokura and N.~Nagaosa, Science {\bf 288} (2000) 462.

\bibitem{Shimomura99PRL} S.~Shimomura, N.~Wakabayashi, H.~Kuwahara and Y.~Tokura,
Phys. Rev. Lett {\bf 83} (1999) 4389.

\bibitem{Burgy01PRL} J.~Burgy, M.~Mayr, V.~Martin-Mayor, A.~Moreo and E.~Dagotto,
Phys. Rev. Lett. {\bf 87} (2001) 277202.

\bibitem{Brink99PRL} J.~v.~d.~Brink, G.~Khaliullin and D.~Khomskii, Phys. Rev. Lett.
{\bf 83} (1999) 5118.

\bibitem{Mizokawa97PRB} T.~Mizokawa and A.~Fujimori, Phys. Rev. B {\bf 56} (1997)
R493.

\bibitem{Y.Waka} Y.~Wakabayashi, Y.~Murakami, I.~Koyama, T.~Kimura, Y.~Tokura, Y.~Moritomo, K.~Hirota and Y.~Endoh: J.Phys.Soc.Jpn. {\bf 69} (2000) 2731.

\bibitem{Y.Waka214} Y.~Wakabayashi, Y.~Murakami, Y.~Moritomo, I.~Koyama,
H.~Nakao, T.~Kiyama, T.~Kimura, Y.~Tokura and N.~Wakabayashi: J. Phys. Soc. Jpn {\bf 70} (2001) 1194.

\bibitem{Murakami214} Y.~Murakami, H.~Kawada, H.~Kawata, M.~Tanaka, T.~Arima,
Y.~Moritomo and Y.~Tokura, Phys. Rev. Let. {\bf 80} (1998) 1932.

\bibitem{Sternlieb96PRL} B.~J.~Sternlieb, J.~P.~Hill, U.~C.~Wildgruber, G.~M. Luke, B.~Nachumi, Y.~Moritomo and Y.~Tokura, Phys. Rev. Lett. {\bf 76} (1996) 2169.

\bibitem{Urushibara95PRB} A.~Urushibara, Y.~Moritomo, T.~Arima, A.~Asamitsu, G.~Kido and Y.~Tokura, Phys. Rev. B {\bf 51} (1995) 14103.


\bibitem{Kubota99JPSJ} M.~Kubota, H.~Yohizawa, Y.~Moritomo, H.~Fujioka, K.~Hirota and Y.~Endoh: J. Phys. Soc. Jpn. {\bf 68} (1999) 2202.

\bibitem{Moritomo96Nature} Y.~Moritomo, A.~Asamitsu, H.~Kuwahara and Y.~Tokura: Nature {\bf 380} (1996) 141.

\bibitem{Kimura98PRB} T.~Kimura, R.~Kumai, Y.~Tokura, J.~W. Li and Y.~Matsui: Phys. Rev. B {\bf 58} (1998) 11081.

\bibitem{Li} J.~Q. Li, Y.~Matsui, T.~Kimura and Y.~Tokura: Phys. Rev. B {\bf 57} (1998) R3205.

\bibitem{Goodenough} J.~B.~Goodenough: Phys. Rev. {\bf 100} (1955) 564.

\bibitem{Argyriou} D.~N.~Argyriou, H.~N.~Bordallo, B.~J.~Campbell, A.~K.~Cheetham, D.~E.~Cox, J.~S.~Gardner, K.~Hanif, A.~dos~Santos and G.~F.~Strouse: Phys. Rev. B {\bf 61} (2000) 15269.

\bibitem{Chatterji} T.~Chatterji, G.~J.~McIntyre, W.~Caliebe, R.~Suryanarayanan, G.~Dhalenne and A.~Revcolevschi: Phys. Rev. B {\bf 61} (2000) 570.

\bibitem{Murakami113} Y.~Murakami, I.~Koyama, M.~Tanaka, H.~Kawata, J.~P. Hill, D.~Gibbs, M.~Blume, T.~Arima, Y.~Tokura, K.~Hirota and Y.~Endoh, Phys. Rev. Lett. {\bf 81} (1998) 582.

\bibitem{Nakamura99PRB} K.~Nakamura, T.~Arima, A.~Nakazawa, Y.~Wakabayashi and Y.~Murakami, Phys. Rev. B {\bf 60} (1999) 2425.

\bibitem{Zimmermann} M.~v.~Zimmermann, J.~P. Hill, D.~Gibbs, M.~Blume, D.~Casa, B.~Keimer,  Y.~Murakami, Y.~Tomioka and Y.~Tokura: Phys. Rev. Lett. {\bf 83} (1999) 4872.

\bibitem{Ishihara1} S.~Ishihara and S.~Maekawa: Phys. Rev. Lett. {\bf 80} (1998) 3799.

\bibitem{Ishihara2} S.~Ishihara and S.~Maekawa: Phys. Rev. B {\bf 58} (1998)13442.

\bibitem{Sasaki_table} S.~Sasaki, KEK Rep. {\bf 88-14}, (1989) 1.

\bibitem{Elfimov} I.~S. Elfimov, V.~I. Anisimov and G.~A. Sawatzky: Phys. Rev. Lett. {\bf 82} (1999) 4264.

\bibitem{Benfatto} M.~Benfatto, Y.~Joly and R.~Natoli: Phys.Rev.Lett. {\bf 83} (1999) 636.

\bibitem{Takahashi} M.~Takahashi, J.~Igarashi and P.~Fulde: J.Phys.Soc.Jpn. {\bf 68} (1999) 2530.

\bibitem{Ishihara_layer} S.~Ishihara and S.~Maekawa: Phys. Rev. B {\bf 62} (2000) 5690.

\bibitem{Merz} M.~Merz, S.~Gerhold, N.~N{\" u}cker, C.~A.~Kuntscher, B.~Burbulla, P.~Schweiss, S.~Schuppler, V.~Chakarian, J.~Freeland, Y.~U.~Idzerda, M.~Kl{\" a}ser, G.~M{\" u}ller-Vogt and Th.~Wolf: Phys. Rev. B {\bf 60} (1999) 9317.

\bibitem{Lee} E.~W.~Lee: Proc. Phys. Soc. {\bf 84} (1964) 693.

\bibitem{Waka_DyZn2} Y.~Wakabayashi, N.~Wakabayashi and T.~Kitai: J. Phys. Soc. Jpn. {\bf 67} (1998) 3901. 


\end{thebibliography}
\end{document}